\documentclass[a4paper,fleqn]{cas-sc}

\usepackage[numbers,sort&compress]{natbib}

\usepackage{amsmath}
\usepackage{amssymb}
\usepackage{bm}
\usepackage{mathtools}
\usepackage{siunitx}
\usepackage{booktabs}
\usepackage{graphicx}
\usepackage{orcidlink}
\let\printorcid\relax 
\let\WriteBookmarks\relax
\ExplSyntaxOn
\ExplSyntaxOff
\begin{document}

\let\WriteBookmarks\relax
\def\floatpagepagefraction{1}
\def\textpagefraction{.001}

\shorttitle{Generalized second-order fluid}
\shortauthors{Marsh and Ewoldt}

\title[mode=title]{Second-order fluids and the choice of objective derivative}

\author[1, 2]{Maxwell C. Marsh\,\orcidlink{0009-0007-2193-5947}}[orcid=0009-0007-2193-5947]

\author[1, 2]{Randy H. Ewoldt\,\orcidlink{0000-0003-2720-9712}}[orcid=0000-0003-2720-9712]
\ead{ewoldt@illinois.edu}
\cormark[1]

\affiliation[1]{
    organization={Department of Mechanical Science and Engineering, Grainger College of Engineering, University of Illinois Urbana-Champaign},
    city={Urbana},
    postcode={61801},
    state={Illinois},
    country={USA}
}
\affiliation[2]{
    organization={Beckman Institute for Advanced Science and Technology, University of Illinois Urbana-Champaign},
    city={Urbana},
    postcode={61801},
    state={Illinois},
    country={USA}
}

\cortext[1]{Corresponding author}

\begin{abstract}
We generalize the second-order fluid constitutive model to accommodate a range of possible frame indifferent time derivatives. 
Using the Gordon--Schowalter derivative, which contains the upper-convected, lower-convected, and corotational (Jaumann-Zaremba) derivatives as subsets, we show how the subjective choice of derivative may change model predictions and we identify features that remain universal independent of this choice. 
Writing these models in terms of generic coefficients is non-universal and may lead to misinterpretations of flow predictions.
However, writing the models in terms of zero-shear material functions or terminal relaxation time reveals universal and robust representations. 
\end{abstract}


\maketitle

\section{Introduction}
\label{intro}
The second-order fluid constitutive model for the extra stress tensor $\boldsymbol{\sigma}$ is typically written as~\cite{bird1987} 
\begin{equation}
\label{eq:SOF}
\boldsymbol{\sigma}
=
\hat{b}_1\boldsymbol{\dot{\gamma}}
+
\hat{b}_2\overset{\triangledown}{\boldsymbol{\dot{\gamma}}}
+
\hat{b}_{11}\boldsymbol{\dot{\gamma}}
\cdot
\boldsymbol{\dot{\gamma}}
\end{equation}
where $\hat{b}_1$, $\hat{b}_2$, and $\hat{b}_{11}$ are constants, $\boldsymbol{\dot{\gamma}} = \nabla\mathbf{v} + (\nabla\mathbf{v})^T$ is the rate of strain tensor, $\nabla\mathbf{v}$ is the velocity gradient tensor with components $(\nabla\mathbf{v})_{ij} = \tfrac{\partial v_j}{\partial x_i}$, and 
\begin{equation}
\label{eq:UC-derivative-gammadot}
\overset{\triangledown}{\boldsymbol{\dot{\gamma}}}
=
\frac{\partial \boldsymbol{\dot{\gamma}}}{\partial t} + (\mathbf{v}\cdot\nabla)\boldsymbol{\dot{\gamma}}
-
(\nabla\mathbf{v})^{T}\cdot\boldsymbol{\dot{\gamma}}
-
\boldsymbol{\dot{\gamma}}\cdot\nabla\mathbf{v}
\end{equation}
is the upper-convected time derivative of the strain rate tensor  (aka Oldroyd's B derivative of the strain rate tensor). 
Equation~\eqref{eq:SOF}, which we call the \textit{upper-convected second-order fluid} (UC-SOF), is considered by many to be a universal description of the asymptotic deviation from Newtonian fluid behavior, akin to a Taylor series expansion about the Newtonian reference state for slow and slowly varying flows.
However, the choice of frame indifferent time derivative is subjective, and there is an infinity of other possible choices.
Therefore, if many other derivatives could instead be used, then how universal is the second-order fluid of eq.~\eqref{eq:SOF}?

In 1955, Rivlin and Ericksen~\cite{Rivlin1955} used the lower-convected derivative (Oldroyd's A derivative) to define recursive kinematic tensors and then in 1960, Coleman and Noll~\cite{Coleman1960} used these kinematic tensors up to second order to define their so-called ``incompressible second-order fluid,'' written as
\begin{equation}
\label{eq:RivErickSOF}
   \boldsymbol{\sigma}
        =
    \alpha \mathbf{A}^{(1)}
+\beta \mathbf{A}^{(1)}\cdot \mathbf{A}^{(1)}
    +\gamma \mathbf{A}^{(2)} 
\end{equation}
where $\alpha$, $\beta$, and $\gamma$ are material constants and $\mathbf{A}^{(1)}$ and $\mathbf{A}^{(2)}$ are known as Rivlin-Ericksen tensors.\footnote[1]{To our knowledge, this is the first use of the term ``second-order fluid''. The terms ``second-grade'' and ``second-order'' fluid were used interchangeably for the following few decades~\cite{Truesdell1965, DunnFosdick1974, McTigue1985} but in recent years ``second-order'' has become more popular in rheological literature. 
}
Converting from this notation, based on the definition of their tensors, we have that $\mathbf{A}^{(1)} = (\nabla\mathbf{v}) + (\nabla\mathbf{v})^{T} = \boldsymbol{\dot\gamma}$ is the rate of strain tensor and $\mathbf{A}^{(2)} = \overset{\vartriangle}{\boldsymbol{\dot\gamma}}$ is the lower-convected derivative of the rate of strain tensor,
\begin{equation} 
\label{eq:LC-derivative-of-strainrate}
\overset{\vartriangle}{\boldsymbol{\dot\gamma}}
=
\frac{\partial \boldsymbol{\dot{\gamma}}}{\partial t} + (\mathbf{v}\cdot\nabla)\boldsymbol{\dot{\gamma}}
+
\nabla \mathbf{v}\cdot\boldsymbol{\dot\gamma}
+
\boldsymbol{\dot\gamma}\cdot(\nabla\mathbf{v})^{T}.
\end{equation}
Thus, we can re-write eq.~\eqref{eq:RivErickSOF} in the notation of \eqref{eq:SOF} as
\begin{equation}
\label{eq:SOF-LC-RivErick}
\boldsymbol{\sigma}
=
\tilde{b}_1\boldsymbol{\dot{\gamma}}
+
\tilde{b}_2\overset{\vartriangle}{\boldsymbol{\dot{\gamma}}}
+
\tilde{b}_{11}\boldsymbol{\dot{\gamma}}
\cdot
\boldsymbol{\dot{\gamma}}
\end{equation}
where $\tilde{b}_1,\tilde{b}_2, \tilde{b}_{11}$ are constants of the \textit{lower-convected second-order fluid} (LC-SOF). 
The use of the lower-convected derivative with Rivlin-Ericksen tensors is not universal. 
While some of the earliest researchers after Rivlin and Ericksen (1960)~\cite{Markovitz1964, Srivastava1963, Ting1963} continued to use the lower-convected derivative, in 1967, Goddard~\cite{Goddard1967} used the co-rotational derivative (aka the Jaumann-Zaremba derivative) to formulate a second-order fluid model, establishing an alternative convention that was subsequently adopted by other authors. 

To date, many researchers have used the lower-convected~\cite{Minale2016, Escott2020, Jitchote2000}, upper-convected~\cite{Kansal2025, Kroo2021, Escott2023}, and co-rotational~\cite{Rallison2012, Subramanian2007, Koch2006, Harnoy1976} derivatives when defining the second-order fluid. 
Considering this diversity of models, Escott and Griffiths~\cite{Escott2023} wrote that the second-order fluid models generated using the three derivatives are ``equivalent so long as the material constants for the model are defined in an appropriate manner.'' 
However, they did not describe how to do this, nor provide any evidence for what differences may (or may not) result.

Here we resolve these open questions 
by considering a family of frame indifferent derivatives, across a spectrum of infinite possibilities, as captured by the Gordon--Schowalter derivative.
We study how predictions change depending on the choice of derivative, based on insight from a range of canonical flows, namely  
homogeneous steady shear, oscillatory shear, steady uniaxial extension, and steady planar extension.

\section{Background}
In 1950, J. G. Oldroyd~\cite{Oldroyd1950} introduced rheologically admissible frame-invariant derivatives for constitutive modeling with his \textit{convected} derivatives.
Today, we refer to them as the \textit{lower-convected} derivative
\begin{equation} 
\label{eq:LC-derivative}
\overset{\vartriangle}{\mathbf{B}}
=
\frac{D\mathbf{B}}{Dt}
+
\nabla \mathbf{v}\cdot\mathbf{B}
+
\mathbf{B}\cdot(\nabla\mathbf{v})^{T}
\end{equation}
and the \textit{upper-convected} derivative
\begin{equation}
\label{eq:UC-derivative}
\overset{\triangledown}{\mathbf{B}}
=
\frac{D\mathbf{B}}{Dt}
-
(\nabla\mathbf{v})^{T}\cdot\mathbf{B}
-
\mathbf{B}\cdot\nabla\mathbf{v}
\end{equation}
where $\boldsymbol{B}$ is an arbitrary symmetric tensor and $\frac{D()}{Dt} = \frac{\partial ()}{\partial t} + (\mathbf{v}\cdot\nabla)() $ is the material derivative operator.
A natural variation of these derivatives is the \textit{corotational derivative}
\begin{equation}
\label{eq:CR-derivative}
\overset{\circ}{\mathbf{B}}
=
\frac{D\mathbf{B}}{Dt}
+
\tfrac{1}{2}\left(\mathbf{\Omega}\cdot
\mathbf{B}
-
\mathbf{B}\cdot\mathbf{\Omega}
\right)
\end{equation}
where $\mathbf{\Omega}=\nabla \mathbf{v} - \left(\nabla \mathbf{v}\right)^{T}$ is a vorticity tensor. This derivative was developed independently by Zaremba in 1903~\cite{Zaremba1903} and by Jaumann in 1911~\cite{Jaumann1911} and so it is referred to as the Jaumann-Zaremba derivative. 

These three derivatives are concisely represented as subsets of the more general Gordon--Schowalter derivative~\cite{Gordon1972,Hinch2021} 
written as
\begin{equation}
\label{eq:GS-derivative}
\overset{\square}{\mathbf{B}}
=
\frac{1+a}{2}
\overset{\triangledown}{\mathbf{B}}
+
\frac{1-a}{2}
\overset{\vartriangle}{\mathbf{B}}
\end{equation}
or conveniently written out as
\begin{equation}
\label{eq:GS-derivative-full}
\overset{\square}{\mathbf{B}}
=
\frac{D\mathbf{B}}{Dt}-
\frac{1+a}{2}
\left[
    (\nabla\mathbf{v})^{T}\cdot\mathbf{B}
    +
    \mathbf{B}\cdot\nabla\mathbf{v}
\right]
+
\frac{1-a}{2}
\left[
\nabla \mathbf{v}\cdot\mathbf{B}
+
\mathbf{B}\cdot(\nabla\mathbf{v})^{T}
\right].
\end{equation}
The choices $a=-1$, $0$, and $1$ recover the lower-convected, corotational, and upper-convected derivatives, respectively~\cite{Larson1988-constitutive}, but any non-integer value for $-1 \le a \le 1$ is equally admissible, covering a spectrum of useful possibilities~\cite{Saramito1995, Hemingway2019}.

\section{Generalized second-order fluid}
We introduce here a \textit{generalized second-order fluid} (G-SOF) constitutive model for the extra stress tensor written in terms of the Gordon--Schowalter derivative. 
For an incompressible fluid, with the extra stress truncated at second order in the velocity gradient,
\begin{equation}
\label{eq:generalized-SOF}
\boldsymbol{\sigma}
=
b_1\dot{\boldsymbol{\gamma}}
+
b_2\overset{\square}{\dot{\boldsymbol{\gamma}}}
+
b_{11}\dot{\boldsymbol{\gamma}}
\cdot
\dot{\boldsymbol{\gamma}},
\end{equation}
where $b_1$, $b_2$, and $b_{11}$ are constants, with a similar meaning to those in eqs.~\eqref{eq:SOF} and \eqref{eq:SOF-LC-RivErick}, and 
\begin{equation}
\label{eq:GS-deriv-gdot}
\overset{\square}{\boldsymbol{\dot{\gamma}}}
=
\frac{D\boldsymbol{\dot{\gamma}}}{Dt}-
\frac{1+a}{2}
\left[
    (\nabla\mathbf{v})^{T}\cdot\boldsymbol{\dot{\gamma}}
    +
    \boldsymbol{\dot{\gamma}}\cdot\nabla\mathbf{v}
\right]
+
\frac{1-a}{2}
\left[
\nabla \mathbf{v}\cdot\boldsymbol{\dot{\gamma}}
+
\boldsymbol{\dot{\gamma}}\cdot(\nabla\mathbf{v})^{T}
\right].
\end{equation}
Equations~\eqref{eq:generalized-SOF}--\eqref{eq:GS-deriv-gdot} fully define the model, with four parameters, $(b_1,b_2,b_{11};a)$. This provides a unifying way to compare the three most common objective derivatives, as well as the spectrum of other admissible choices within the Gordon--Schowalter family.
This allows us to study how model predictions depend on the four model parameters, and in particular $a$, which represents the choice of time derivative.

\subsection{Steady simple shear}
\begin{figure}
    \centering
    \includegraphics[width=0.5\textwidth]{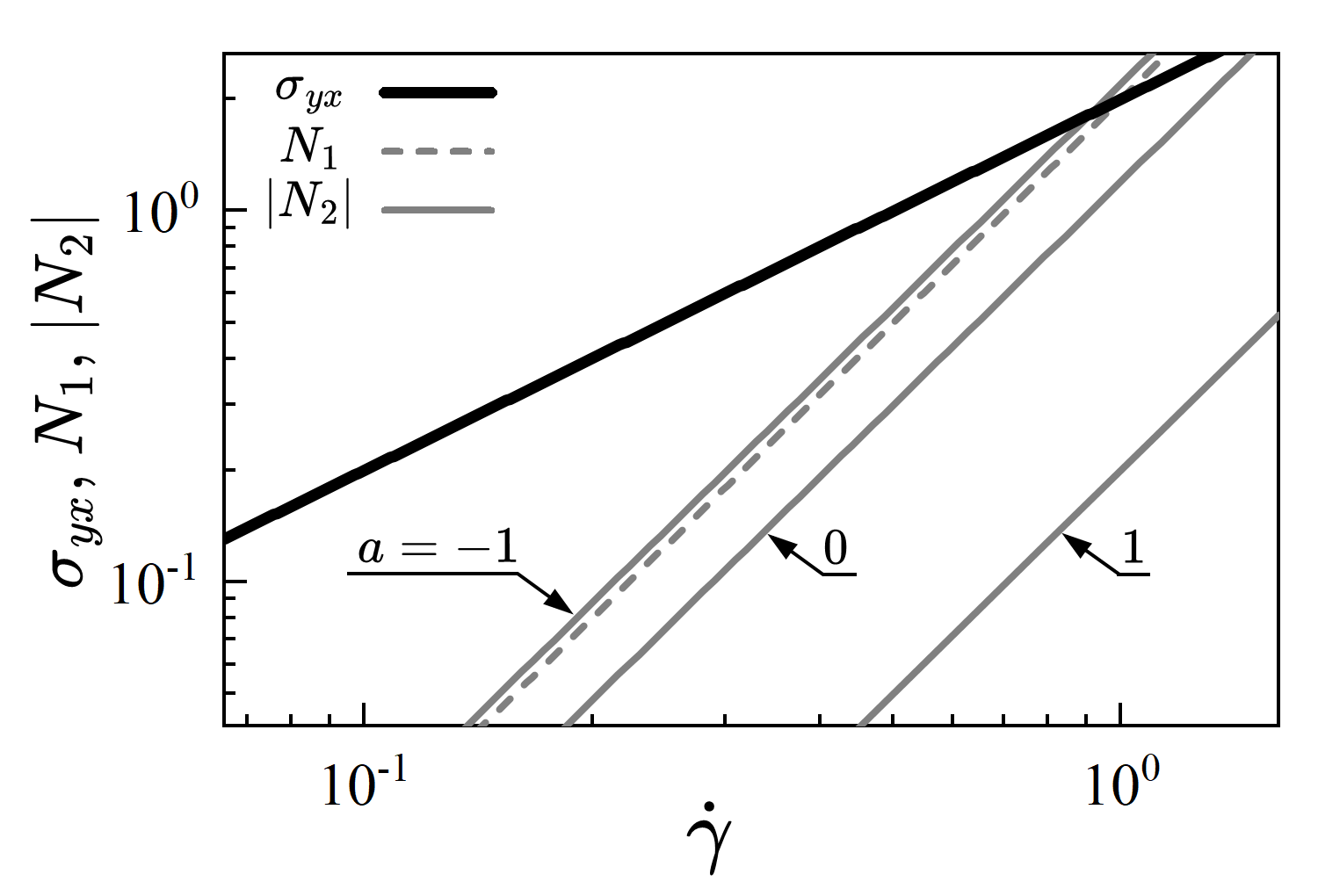}
    \caption{
    Stress predictions for homogeneous steady shear using three different time derivatives in the SOF with fixed generic coefficients: lower-convected ($a=-1$), corotational ($a=0$), and upper-convected ($b=1$). 
    Nominal parameters used are $b_1=2$~Pa.s, $b_2=-1$~Pa.s$^2$, and $b_{11}=-0.2$~Pa.s$^2$. 
    The shear stress $\sigma_{yx}$ and first normal stress difference $N_1$ do not depend on $a$, but $N_2$ does and its magnitude changes significantly. 
    \label{fig:steady-shear-stress}
    }
\end{figure}

We first consider homogeneous steady simple shear with velocity field $\mathbf{v}=\left(\dot{\gamma}y,0,0\right)$. Using this velocity field in eq.~\eqref{eq:GS-deriv-gdot} gives
\begin{equation}
\label{eq:GS-rate-of-strain}
\overset{\square}{\boldsymbol{\dot{\gamma}}}
=
\begin{bmatrix}
-(1+a)\dot{\gamma}^{2} & \frac{d\dot{\gamma}}{dt} & 0 \\
\frac{d\dot{\gamma}}{dt} & (1-a)\dot{\gamma}^{2} & 0 \\
0 & 0 & 0
\end{bmatrix}
\end{equation}
which we substitute into eq.~\eqref{eq:generalized-SOF} to get the extra stress,
\begin{equation}
\label{eq:G-SOF-stress}
\boldsymbol{\sigma}
=
\begin{bmatrix}
\left[b_{11}-(1+a)b_2\right]\dot{\gamma}^{2}
&
b_1\dot{\gamma}+b_2\frac{d\dot{\gamma}}{dt}
&
0
\\
b_1\dot{\gamma}+b_2\frac{d\dot{\gamma}}{dt}
&
\left[b_{11}+(1-a)b_2\right]\dot{\gamma}^{2}
&
0
\\
0 & 0 & 0
\end{bmatrix}.
\end{equation}
For steady simple shear, $\frac{d\dot{\gamma}}{dt}=0$. 
The predicted response, in terms of steady shear material functions 
\begin{equation}
\label{eq:matl-func-def}
\eta
=
\frac{\sigma_{yx}}{\dot{\gamma}},
\qquad
\Psi_1
= \frac{N_1}{\dot\gamma^2}=
\frac{\sigma_{xx}-\sigma_{yy}}{\dot{\gamma}^{2}} ,
\qquad
\Psi_2
= \frac{N_2}{\dot\gamma^2} =
\frac{\sigma_{yy}-\sigma_{zz}}{\dot{\gamma}^{2}}, 
\end{equation}
is 
\begin{equation}
    \label{SS-matl-funcs}
    \begin{aligned}
        \eta &= b_1 \\ 
        \Psi_1 &= -2b_2 \\
        \Psi_2 &= b_{11}+(1-a)b_2        
    \end{aligned}
\end{equation}
for the G-SOF. 
The results from eq.~\eqref{SS-matl-funcs} are plotted in Figure~\ref{fig:steady-shear-stress} in terms of $\sigma_{yx}$, $N_1$, and $N_2$, which demonstrates the non-universality of this model. By fixing the $b$ parameters, the predictions for $\sigma_{yx}$ and $N_1$ remain independent of $a$, but $|N_2|$ varies, taking values both greater than and less than $N_1$.  The product $ab_2$ appears within $\Psi_2$, which can be rationalized by expanding the Gordon-Schowalter derivative in eq.~\eqref{eq:GS-deriv-gdot} and gathering terms (Appendix~A).

The G-SOF predicts constant steady shear material functions, which we interpret as the zero shear values, denoted as $\eta=\eta_0$, $\Psi_1 = \Psi_{1,0}$, and $\Psi_2 =\Psi_{2,0}$.
With this notation, we can solve eq.~\eqref{SS-matl-funcs} for the $b$ coefficients as a function of the zero-shear material functions,
\begin{equation}
\label{eq:bs-as-coeffs}
    \begin{aligned}
        b_1 &= \eta_0  \\ 
        b_2 &= -\frac{1}{2}\Psi_{1,0} \\
        b_{11} &= \Psi_{2,0}+\frac{1-a}{2}\Psi_{1,0}.
    \end{aligned}    
\end{equation}
From eq.~\eqref{eq:bs-as-coeffs}, 
we can then rewrite the G-SOF, eq.~\eqref{eq:generalized-SOF}, in terms of the zero-shear material functions as 
\begin{equation}
\label{gen-SOF-matl-func}
    \boldsymbol{\sigma}
=
\eta_0\boldsymbol{\dot{\gamma}}
-
\frac{\Psi_{1,0}}{2}\overset{\square}{\boldsymbol{\dot{\gamma}}}
+
\left(\Psi_{2,0}
+
\frac{1-a}{2}\Psi_{1,0} \right)\boldsymbol{\dot{\gamma}}
\cdot
\boldsymbol{\dot{\gamma}}.
\end{equation}
This is a key result: eq.~\eqref{gen-SOF-matl-func} will consistently predict the zero-shear material functions as $\eta=\eta_0$, $\Psi_1 = \Psi_{1,0}$, and $\Psi_2 = \Psi_{2,0}$, independent of the value of $a$, that is, independent of the choice of time derivative within the Gordon--Schowalter family. 

\subsection{Oscillatory shear}
\label{sec-oscillatory}
Oscillatory shear will reveal a relaxation time, $\tau$. 
For an oscillatory strain input $\gamma(t)=\gamma_0\sin(\omega t)$, the resulting strain rate is $\dot{\gamma}(t) = \dot{\gamma}_0\cos(\omega t) = \gamma_0\omega\cos(\omega t)$, 
and its derivative 
$\tfrac{d\dot{\gamma}(t)}{dt} = -\gamma_0\omega^2\sin(\omega t)$.
Substituting into eq.~\eqref{eq:G-SOF-stress} gives the stress components
\begin{subequations} \label{eq:SOF-SAOS-stresses}
\begin{align}
\sigma_{yx}(t) &= \gamma_0\left(
-b_2\omega^2 \sin(\omega t)
+b_1\omega \cos(\omega t)
\right) \\
N_1(t) &= \dot\gamma_0^2 \left( -b_2\left[1+\cos(2\omega t)\right] \right) \label{eq:SOF-SAOS-N1}\\
N_2(t) &= \dot\gamma_0^2 \tfrac{1}{2}\left[b_{11} + \left(1-a\right)b_2 \right]\left[1+\cos(2\omega t)\right] \label{eq:SOF-SAOS-N2}
\end{align}
\end{subequations}
where we use $\dot\gamma^2=\dot{\gamma}_0^2\cos^2(\omega t)=\frac{\dot{\gamma}_0^2}{2}\left[1+\cos(2\omega t)\right]$.
Notice that $b_2$ appears in each stress component and $a$ appears only in $N_2(t)$.
In terms of viscoelastic moduli, defined by the form $\sigma_{yx} = \gamma_0\left[
G'\sin(\omega t) + G'' \cos(\omega t)\right]$, we have
\begin{subequations}
\label{eq:SOF-Gp-and-Gpp}
   \begin{align}
G'(\omega)
&=
-b_2\omega^2
\label{SOF-Gp-bs} \\
G''(\omega)
&=
b_1\omega
\label{SOF-Gpp-bs}
\end{align} 
\end{subequations}
which are not affected by $a$. 
The dynamic normal stress coefficients can be defined by the form~\cite{Ferry1980}
\begin{subequations}
\label{ferry-def}
    \begin{align}
    N_1(t)  &= \dot\gamma_0^2 \left[
    \Psi_1^{d}
    +
    \Psi_1^{\prime}\cos (2\omega t)
    -
    \Psi_1^{\prime\prime}\sin (2\omega t) \right] \\
    N_2(t)  &= \dot\gamma_0^2 \left[
    \Psi_2^{d}
    +
    \Psi_2^{\prime}\cos (2\omega t)
    -
    \Psi_2^{\prime\prime}\sin (2\omega t) \right]
    \end{align}
\end{subequations}
where $\Psi_i^d$ denotes the constant offset normal stress component of the $i$th normal stress coefficient. 
Comparing this to eqs.~\eqref{eq:SOF-SAOS-N1}--\eqref{eq:SOF-SAOS-N2} we have
\begin{subequations}
\label{eq:oscillatory-Psi's}
   \begin{align}
    \label{eq:psi1-def}
        \Psi_1^{d}&=\Psi_1^{\prime}= -b_2, \quad \Psi''_1 =0\\
        \Psi_2^{d}&=\Psi_2'=\tfrac{1}{2}[b_{11}+\left(1-a\right)b_2]     \label{eq:psi2-def}, \quad \Psi''_2 =0.
    \end{align} 
\end{subequations}
\begin{figure}
    \centering
    \includegraphics[width=0.5\textwidth]{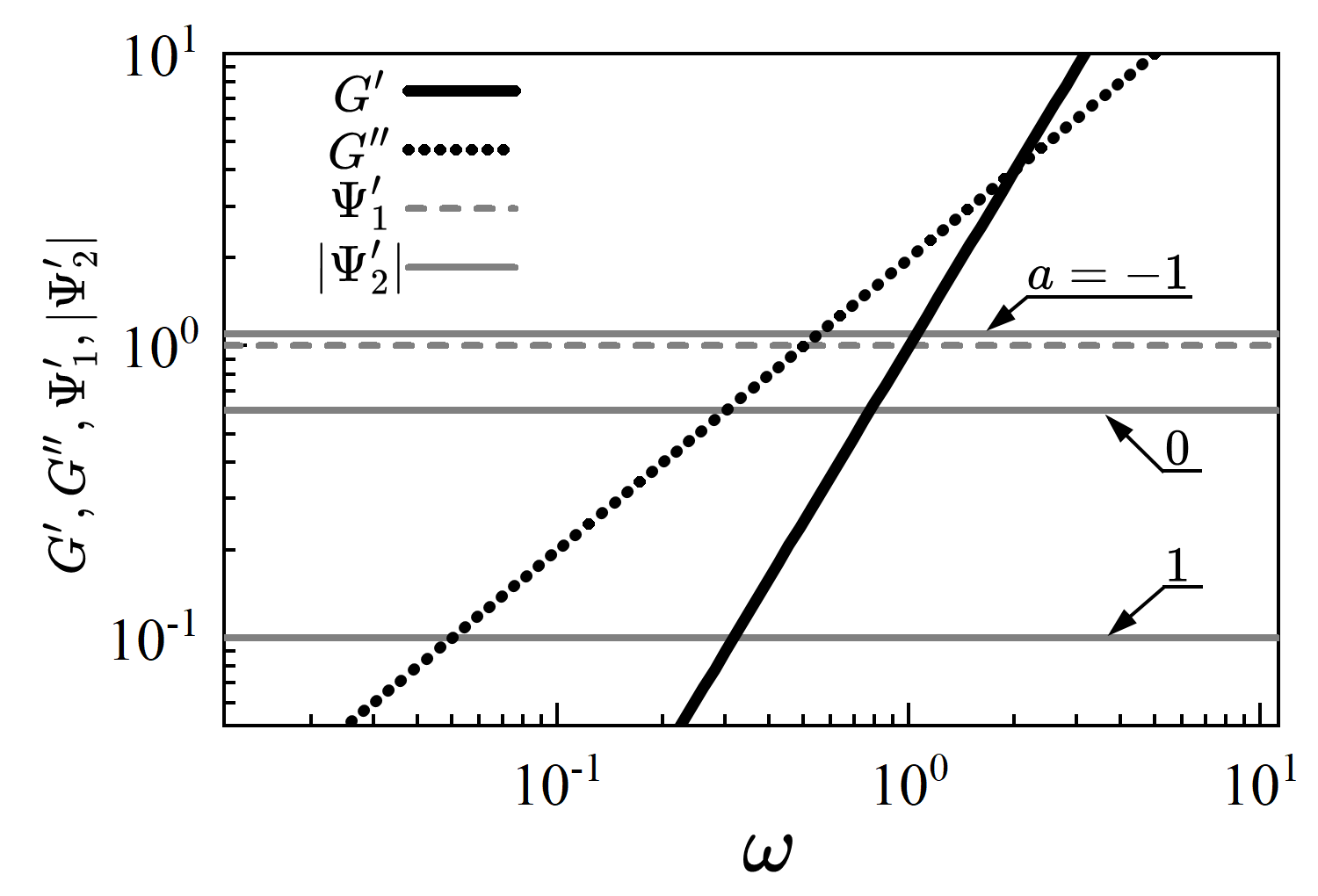}
    \caption{
    Small amplitude oscillatory shear predictions of the SOF using the same three time derivatives and generic $b$ parameters as in Figure~\ref{fig:steady-shear-stress}. 
    The dynamic moduli and first normal stress difference coefficient ($\Psi_1^d=\Psi_1' $) do not depend on $a$ but the magnitude of the second normal stress difference coefficient ($\Psi_2^d=\Psi_2' $) does and it varies significantly. 
    }
    \label{fig:osc-shear-stress}
\end{figure}

Figure~\ref{fig:osc-shear-stress} plots the oscillatory shear material functions 
versus frequency using the same parameters as in Figure~\ref{fig:steady-shear-stress}. 
Like in the steady shear example, only the second normal stress component depends on the choice of derivative while the dynamic moduli and the first normal stress component remain unchanged.

One consequence of this behavior is the robustness of the relaxation time to choice of derivative. From the terminal regime definition of the relaxation time, we have
\begin{equation}
\tau
\equiv
\lim_{\omega\rightarrow0}
\frac{G'(\omega)}
{\omega G''(\omega)}
=
-\frac{b_2}{b_1}.
\label{relax-time-bs}
\end{equation}
This is independent of $a$, a significant preserved prediction for G-SOF across any choice of time derivative. 
In terms of zero-shear material functions, substituting from eq.~\eqref{eq:bs-as-coeffs},
this characteristic relaxation time is 
\begin{equation}
    \tau=\frac{\Psi_{1,0}}{2\eta_0}
\label{relax-time-matl-funcs}
\end{equation}
and therefore the interrelation $\Psi_{1,0} = 2\eta_0\tau$, which anticipates nonlinear behavior from linear behavior, is universal across any choice of time derivative within the G-SOF framework. 

We can rewrite the G-SOF oscillatory shear material functions in terms of $\tau$ and zero-shear material functions. 
For example, eq.~\eqref{SOF-Gp-bs} becomes $G'=\tfrac{\eta_0}{\tau}(\tau \omega)^2=\tfrac{\eta_0}{\tau}De^2$ where $De=\tau\omega$ is a Deborah number. 
Moreover, if we substitute eq.~\eqref{eq:bs-as-coeffs} into eqs.~\eqref{eq:SOF-Gp-and-Gpp} and eq.~\eqref{eq:oscillatory-Psi's}, we see
\begin{subequations}
    \label{matl-func-SAOS}
    \begin{align}
        G'(\omega)
        &=
        \tfrac{1}{2}\Psi_{1,0}
        \omega^2
        \label{SOF-Gp-matl-func} \\
        G''(\omega)
        &=
        \eta_0\omega\\
        \Psi_1^d(\omega)=\Psi_1'(\omega) 
        &= \tfrac{1}{2}\Psi_{1,0}\\
        \Psi_2^d(\omega)=\Psi_2'(\omega) 
        &= \tfrac{1}{2}\Psi_{2,0}    
    \end{align}
\end{subequations}
revealing an $a$-independent interrelation between the oscillatory shear and steady shear material functions for the G-SOF, even for the second normal stress difference coefficients.
The G-SOF governing equation can also be rewritten in terms of $\tau$ (Appendix~B).

\subsection{Uniaxial extension}
\begin{figure}
    \centering
    \includegraphics[width=0.5\textwidth]{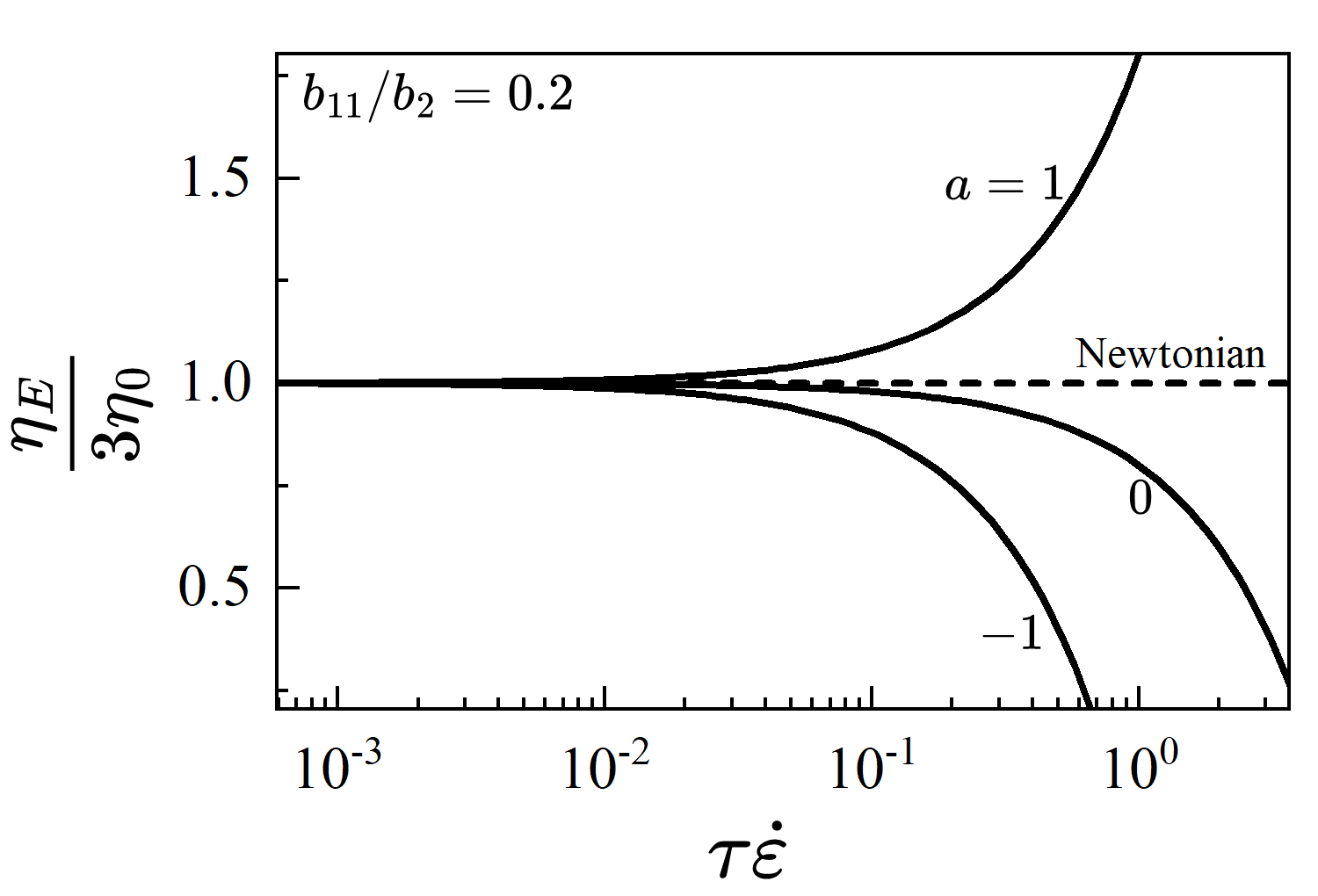}
    \caption{
    SOF prediction of uniaxial extensional viscosity as a function of Weissenberg number $Wi=\tau\dot\varepsilon$, eq.~\eqref{eq:dimless-ext-visc}, with fixed generic coefficient ratio $b_{11}/b_2=0.2$ as a representative example. 
    Qualitatively different behavior is predicted depending on the choice of derivative, changing from extensional thickening ($a=1$) to extensional thinning ($a=0$, $a=-1$).
    This problem is resolved by the zero-shear material function representation of eq.~\eqref{ext-visc-matl-funcs} or eq.~\eqref{ext-visc-matl-funcs-tau}.
    \label{fig:steady-extensional-stress}
    }
\end{figure}

In extension, the choice of derivative can fundamentally change the type of nonlinearity instead of just changing the constant prefactor.
In homogeneous uniaxial extension with velocity field $\mathbf{v}=(
\dot{\varepsilon}x,
-\frac{1}{2}\dot{\varepsilon}y,
-\frac{1}{2}\dot{\varepsilon}z
)$,
eq.~\eqref{eq:GS-deriv-gdot} gives
\begin{equation}
\overset{\square}{\boldsymbol{\dot{\gamma}}}
=
\begin{bmatrix}
2\frac{d\dot{\varepsilon}}{dt}
-4a\dot{\varepsilon}^{2}
& 0 & 0
\\
0 &
-\frac{d\dot{\varepsilon}}{dt}
-a\dot{\varepsilon}^{2}
& 0
\\
0 & 0 &
-\frac{d\dot{\varepsilon}}{dt}
-a\dot{\varepsilon}^{2}
\end{bmatrix}
\label{eq:uni-ext-gdot-tensor}
\end{equation}
which we can substitute into eq.~\eqref{eq:generalized-SOF} resulting in
\begin{equation}
\boldsymbol{\sigma}
=
\begin{bmatrix}
2b_1\dot{\varepsilon}
+2b_2\frac{d\dot{\varepsilon}}{dt}
+4(b_{11}-ab_2)\dot{\varepsilon}^{2}
& 0 & 0
\\
0 &
-b_1\dot{\varepsilon}
-b_2\frac{d\dot{\varepsilon}}{dt}
+(b_{11}-ab_2)\dot{\varepsilon}^{2}
& 0
\\
0 & 0 &
-b_1\dot{\varepsilon}
-b_2\frac{d\dot{\varepsilon}}{dt}
+(b_{11}-ab_2)\dot{\varepsilon}^{2}
\end{bmatrix}.
\label{G-SOF-uni-ext-stress}
\end{equation}
For steady flow, 
the uniaxial extensional viscosity
is then
\begin{equation}
    \label{eq:dimless-ext-visc}
    \eta_E
    \equiv
    \frac{\sigma_{xx}-\sigma_{yy}}{\dot{\varepsilon}}=
    3b_1
    +
    3(b_{11}-ab_2)\dot{\varepsilon}
    \qquad \text{or} \qquad
    \frac{\eta_E}{3\eta_0}=1+\left(a-\frac{b_{11}}{b_2}\right)\tau\dot\varepsilon,
\end{equation}
using $b_1=\eta_0$ from eq.~\eqref{eq:bs-as-coeffs}, $b_2=-\tau b_1$ from eq.~\eqref{relax-time-bs}, and constructing the dimensionless strain rate as $\tau\dot\varepsilon$. 
The result of eq.~\eqref{eq:dimless-ext-visc} is plotted in Figure~\ref{fig:steady-extensional-stress} using the same model parameters from Figures~\ref{fig:steady-shear-stress} and \ref{fig:osc-shear-stress} to show fundamentally different nonlinear behavior depending on derivative choice. 
With these fixed $b$ parameters, the upper-convected derivative predicts extensional thickening whereas the corotational and lower-convected derivatives predict extensional thinning. 

We can universalize the prediction for $\eta_E$ by replacing the $b$ parameters with universally-defined quantities, such as zero-shear material functions or terminal relaxation time $\tau$. 
For example, substituting eq.~\eqref{eq:bs-as-coeffs} into eq.~\eqref{eq:dimless-ext-visc}, we find that the $a$-dependence cancels out, giving
\begin{equation}
\eta_E
=
3\eta_0
+
\frac{3}{2}
\left(
\Psi_{1,0}+2\Psi_{2,0}
\right)\dot{\varepsilon}
\label{ext-visc-matl-funcs}
\end{equation}
or in terms of a Weissenberg number, $Wi=\tau\dot\varepsilon$, this is 
\begin{equation}
\frac{\eta_E}{3\eta_0}
= 1 + \left( 1 + \frac{2\Psi_{2,0}}{\Psi_{1,0}}
\right)\tau\dot{\varepsilon}.
\label{ext-visc-matl-funcs-tau}
\end{equation}
Equations~\eqref{ext-visc-matl-funcs} and \eqref{ext-visc-matl-funcs-tau} are universal predictions of the G-SOF independent of the choice of $a$.

\subsection{Planar extension}
Lastly, in planar extension with velocity field $\mathbf{v}=\left(\dot{\varepsilon}_Px,-\dot{\varepsilon}_Py,0\right)$ the extra stress tensor from eq.~\eqref{eq:generalized-SOF} is
\begin{equation}
\boldsymbol{\sigma}
=
\begin{bmatrix}
2b_1\dot{\varepsilon}_P
+2b_2\frac{d\dot{\varepsilon}_P}{dt}
+4(b_{11}-ab_2)\dot{\varepsilon}_P^{2}
& 0 & 0
\\
0 &
-2b_1\dot{\varepsilon}_P
-2b_2\frac{d\dot{\varepsilon}_P}{dt}
+4(b_{11}-ab_2)\dot{\varepsilon}_P^{2}
& 0
\\
0 & 0 & 0
\end{bmatrix}.
\label{planar-ext-stress}
\end{equation}
For steady flow, the first and second planar extensional viscosities are then
\begin{align}
\label{eq:planar-visc-1}
\eta_{P1}
&\equiv
\frac{\sigma_{xx}-\sigma_{yy}}{\dot{\varepsilon}_P}
=4b_1
\\
\eta_{P2}
&\equiv
\frac{\sigma_{zz}-\sigma_{yy}}{\dot{\varepsilon}_P}
=2b_1-4\left(b_{11}-ab_2\right)\dot{\varepsilon}_P. 
\label{eq:planar-visc-2}
\end{align}
Taking $b_1=\eta_0$ and using $\tau$ to construct a dimensionless strain rate, this is $\eta_{P1}=4\eta_0$ and $\tfrac{\eta_{P2}}{2\eta_0}=1-2
\left(
a-\tfrac{b_{11}}{b_2}
\right)
\tau\dot\varepsilon_P$.
We see that $\eta_{P1}$ is constant up to second order, in contrast to the uniaxial extensional viscosity $\eta_E$, eq.~\eqref{eq:dimless-ext-visc}. 
This is consistent with experimental results, for example LDPE shows a weaker and delayed planar extensional thickening of $\eta_{P1}$ compared to its uniaxial extensional thickening of $\eta_E$~\cite{Zatloukal2016}.
However, we see that $\eta_{P2}$ does show $\dot\varepsilon$- and $a$-dependence, eq.~\eqref{eq:planar-visc-2}. The functional form is similar to that of $\eta_E$. 

We can again find that the $a$ dependence is removed by replacing the $b$ parameters with zero-shear material functions and relaxation time,
\begin{subequations}
\label{matl-func-planar-ext}
    \begin{align}
        \eta_{P1}
        &=4\eta_0\\
        \frac{\eta_{P2}}{2\eta_0}
        &= 1 - 2\left( 1 + \frac{2\Psi_{2,0}}{\Psi_{1,0}}
        \right)\tau\dot{\varepsilon}_P,  
    \end{align}
\end{subequations}
where $Wi=\tau\dot\varepsilon_P$. 
In this way, we can compare the weakly non-Newtonian behavior across all four flow scenarios~\cite{King2026b, Daivis2007}. For example, as a function of $Wi$, $\eta_{P2}$ changes twice as rapidly as $\eta_E$, eq.~\eqref{ext-visc-matl-funcs-tau}, and both depend on the ratio $\Psi_{2,0}/\Psi_{1,0}$.

\section{Conclusions}

We have shown that stress predictions of the incompressible second-order fluid can depend on the choice of time derivative. 
This is especially notable in predictions of $N_2$, $\eta_E$, and $\eta_{P2}$. 
Some predictions are universal, such as the terminal relaxation time $\tau$ and the interrelation $\Psi_{1,0}=2\eta_0\tau$, which result for any choice of time derivative within the Gordon-Schowalter family. 
For the non-universal predictions, we have shown that each can be made universal by replacing the generic $b$ coefficients with zero-shear material functions, namely $\eta_0$, $\Psi_{1,0}$, $\Psi_{2,0}$, or the terminal relaxation time $\tau$. 

The generic coefficient non-universal predictions are important for fluid mechanics analysis. 
The second-order fluid model was developed to predict a rigorous and universal first deviation from Newtonian behavior in the limit of $De=0^+$ and $Wi=0^+$. 
It has been applied to a wide range of complex fluid dynamics scenarios, several of which are governed by streamline tension in shear or extension, which may correlate with $N_2$, $\eta_E$, or $\eta_{P2}$. 
This includes particle focusing in pressure-driven microfluidic flows~\cite{Ho1976, Kim2019}, edge fracture in rotational shear rheometry~\cite{Tanner1983, Keentok1999, Hemingway2019}, rod-climbing~\cite{Bonn2004, Weissenberg1947} or rod-dipping~\cite{Boyer2011} in a bath of fluid, free surface bulging in open-channel flow~\cite{Dai2014, Tanner1970}, and other flows involving $N_2$~\cite{Maklad2021, Datta2022, Boulogne2012}. 
Extensional kinematics appear in application to flow around rigid spheres~\cite{Apte2025} and particle migration in a microchannel crosslot extensional flow~\cite{Hsiao2017}. 
Any results presented in terms of generic coefficients may be non-universal, depending on the choice of frame indifferent derivative.
However, results presented in terms of zero-shear material functions and terminal relaxation time should be universal across the Gordon-Schowalter family of derivatives.

The G-SOF, represented in terms of zero-shear material functions, eq.~\eqref{gen-SOF-matl-func}, is a key result that would make universal predictions. 
Equation~\eqref{gen-SOF-matl-func} depends on $a$, but this is removed for any specific choice of it, which has historically been the case. 
For example, in terms of the lower-convected, corotational, or upper-convected derivative, eq.~\eqref{gen-SOF-matl-func} reduces to
\begin{align}
        \label{eq:equivalent-SOF-LC}
        \boldsymbol{\sigma}
        &=
        \eta_0\boldsymbol{\dot{\gamma}}
        -\tfrac{1}{2}\Psi_{1,0}
        \overset{\vartriangle}{\boldsymbol{\dot{\gamma}}}
        +
        \left(\Psi_{2,0}
        +
        \Psi_{1,0} \right)\boldsymbol{\dot{\gamma}}
        \cdot
        \boldsymbol{\dot{\gamma}}\\      
        \label{eq:equivalent-SOF-CR}
        \boldsymbol{\sigma}
        &=
        \eta_0\boldsymbol{\dot{\gamma}}
        -\tfrac{1}{2}\Psi_{1,0}
        \overset{\circ}{\boldsymbol{\dot{\gamma}}}
        +
        \left(\Psi_{2,0}
        +
        \tfrac{1}{2}\Psi_{1,0} \right)\boldsymbol{\dot{\gamma}}
        \cdot
        \boldsymbol{\dot{\gamma}}\\
        \label{eq:equivalent-SOF-UC}
        \boldsymbol{\sigma}
        &=
        \eta_0\boldsymbol{\dot{\gamma}}
        -\tfrac{1}{2}\Psi_{1,0}
        \overset{\triangledown}{\boldsymbol{\dot{\gamma}}}
        +
        \Psi_{2,0}
        \boldsymbol{\dot{\gamma}}
        \cdot
        \boldsymbol{\dot{\gamma}}.
\end{align}
In terms of Rivlin-Ericksen tensors, eq.~\eqref{eq:RivErickSOF}, equivalent to the lower-convected form, eq.~\eqref{eq:SOF-LC-RivErick}, the result is, 
\begin{equation}
    \label{general-riv-erick-SOF}
        \boldsymbol{\sigma}
        =
        \eta_0\mathbf{A}^{(1)}
        -\tfrac{1}{2}\Psi_{1,0}
        \mathbf{A}^{(2)}
        +
        \left(\Psi_{2,0}
        +
        \Psi_{1,0} \right)\mathbf{A}^{(1)}
        \cdot
        \mathbf{A}^{(1)},    
\end{equation}
where $\mathbf{A}^{(1)}$ and $\mathbf{A}^{(2)}$ have the same meaning as in Section~\ref{intro}.
When dealing with results that have used generic coefficients, such as those in prior work, these can be interpreted by converting to the universal zero-shear material function representation.

When choosing a universal representation of the second order fluid, from the options in eqs.~\eqref{eq:equivalent-SOF-LC}--\eqref{general-riv-erick-SOF}, the specific choice of derivative may be motivated by mathematical convenience, its relation to microstructural motivations, or its relation to other nonlinear viscoelastic constitutive models of interest. 
The upper-convected form in eq.~\eqref{eq:equivalent-SOF-UC} is a convenient choice due to its applicability to bead-spring systems and the fact that the coefficient on the $\boldsymbol{\dot{\gamma}}^2$ term is exclusively $\Psi_{2,0}$, 
but other representations can also be equivalent if the correct representation is used.

\section*{CRediT authorship contribution statement}
\textbf{Maxwell C. Marsh}: Conceptualization, Formal analysis, Investigation, Writing – original draft, Writing – review and editing. \textbf{Randy H. Ewoldt}: Conceptualization, Formal analysis, Investigation, Supervision, Writing – original draft, Writing – review and editing.

\section*{Acknowledgments}
The authors acknowledge support for this work as part of the Regenerative Energy Efficient Manufacturing of Thermoset Polymeric Materials (REMAT), an Energy Frontier Research Center funded by the U.S. Department of Energy, Office of Science, Basic Energy Sciences at the University of Illinois Urbana-Champaign under Award No. DE-SC0023457.

\appendix
\section*{Appendix A. Gordon--Schowalter representation}
\label{appendix-A}
We can expand the Gordon--Schowalter derivative of the rate of strain tensor in eq.~\eqref{eq:GS-deriv-gdot} as
\begin{equation}
\label{AppendixA-1}
    \begin{aligned}
    \overset{\square}{\boldsymbol{\dot{\gamma}}}
    =
    \frac{D\boldsymbol{\dot{\gamma}}}{Dt}
    &-\frac{1}{2}
    \left[(
    \nabla\mathbf{v})^{T}
    \cdot
    \boldsymbol{\dot{\gamma}}
    +\boldsymbol{\dot{\gamma}}
    \cdot
    (\nabla\mathbf{v})
    -(\nabla\mathbf{v})
    \cdot\boldsymbol{\dot{\gamma}}
    -\boldsymbol{\dot{\gamma}}
    \cdot(\nabla\mathbf{v})^{T}
    \right]\\
        &-\frac{a}{2}
    \left[(
    \nabla\mathbf{v})^{T}
    \cdot
    \boldsymbol{\dot{\gamma}}
    +\boldsymbol{\dot{\gamma}}
    \cdot
    (\nabla\mathbf{v})
    +(\nabla\mathbf{v})
    \cdot\boldsymbol{\dot{\gamma}}
    +\boldsymbol{\dot{\gamma}}
    \cdot(\nabla\mathbf{v})^{T}
    \right].
\end{aligned} 
\end{equation}
Grouping like terms 
on the RHS gives
\begin{equation}
\label{AppendixA-2}
\begin{aligned}
    \overset{\square}{\boldsymbol{\dot{\gamma}}}
    =
    \frac{D\boldsymbol{\dot{\gamma}}}{Dt}
    &-\frac{1}{2}
    \left[-\left(
    (\nabla\mathbf{v})
    -(\nabla\mathbf{v})^{T}
    \right)
    \cdot\boldsymbol{\dot{\gamma}}
    +
    \boldsymbol{\dot{\gamma}}\cdot\left(
    (\nabla\mathbf{v})
    -(\nabla\mathbf{v})^{T}
    \right)
    \right]\\
        &-\frac{a}{2}
    \left[\left(
    (\nabla\mathbf{v})^{T}
    +(\nabla\mathbf{v})
    \right)
    \cdot\boldsymbol{\dot{\gamma}}
    +
    \boldsymbol{\dot{\gamma}}\cdot\left(
    (\nabla\mathbf{v})^{T}
    +(\nabla\mathbf{v})
    \right)
    \right].
\end{aligned} 
\end{equation}
We recognize that, by definition, $\boldsymbol{\dot{\gamma}} = (\nabla\mathbf{v})^{T} + (\nabla\mathbf{v})$ and  
$\mathbf{\Omega}=\nabla \mathbf{v} - \left(\nabla \mathbf{v}\right)^{T}$, and so we can rewrite eq.~\eqref{AppendixA-2} as
\begin{equation}
\label{AppendixA-3}
    \overset{\square}{\boldsymbol{\dot{\gamma}}}
    =
    \frac{D\boldsymbol{\dot{\gamma}}}{Dt}
    +\frac{1}{2}
    \left[
    \boldsymbol{\Omega}\cdot\boldsymbol{\dot{\gamma}}
    -\boldsymbol{\dot{\gamma}}\cdot\boldsymbol{\Omega}
    \right]
    -a\boldsymbol{\dot{\gamma}}^2.
\end{equation}
The first and second terms on the RHS are exactly the corotational derivative, eq.~\eqref{eq:CR-derivative}, and so we can write eq.~\eqref{AppendixA-3}
\begin{equation}
    \label{AppendixA-4}
    \overset{\square}{\boldsymbol{\dot{\gamma}}}
    =
    \overset{\circ}{\boldsymbol{\dot{\gamma}}}-a\boldsymbol{\dot{\gamma}}^2.
\end{equation}
(The corresponding result for an arbitrary symmetric tensor is $\overset{\square}{\mathbf{B}} =  \overset{\circ}{\mathbf{B}}-\frac{a}{2}\left[\mathbf{B}\cdot\boldsymbol{\dot{\gamma}}+\boldsymbol{\dot{\gamma}}\cdot\mathbf{B}\right]$, following a similar derivation.) The generalized second-order-fluid model of eq.~\eqref{eq:generalized-SOF} can therefore equivalently be written as
\begin{equation}
\label{generalized_SOF_alt}
\boldsymbol{\sigma}
=
b_1\dot{\boldsymbol{\gamma}}
+
b_2\overset{\circ}{\dot{\boldsymbol{\gamma}}}
+
\left(b_{11}-ab_2\right)\dot{\boldsymbol{\gamma}}^2.
\end{equation}
In this way, the $a$ term is preserved and we see why it is naturally grouped as $ab_2$ and why it influences predictions associated with the $\boldsymbol{\dot\gamma}^2$ term, for example as in the results of eqs.~\eqref{SS-matl-funcs}, \eqref{eq:psi2-def}, \eqref{eq:dimless-ext-visc}, and \eqref{eq:planar-visc-2}.

\section*{Appendix B. Alternative G-SOF form}
\label{Appendix-Altenrative G- SOF}
Analogous to eq.~\eqref{gen-SOF-matl-func}, we can choose other preserved quantities to write the G-SOF. 
For example, in terms of $\eta_0$, $\tau$, and $\epsilon\equiv \Psi_{2,0}/\Psi_{1,0}$, the $b$ coefficients of eq.~\eqref{eq:generalized-SOF} can be solved from eqs.~\eqref{eq:bs-as-coeffs} and \eqref{relax-time-matl-funcs} to give
\begin{equation}
\label{AppendixC-3}
\begin{aligned}
        b_1 &= \eta_0  \\ 
        b_2 &= -\eta_0\tau \\
        b_{11} &= \eta_0\tau \left( 2\epsilon + 1-a\right).
\end{aligned}
\end{equation}
Substituting into eq.~\eqref{eq:generalized-SOF} gives
\begin{equation}
\label{gen-SOF-matl-func-tau}
    \boldsymbol{\sigma}
=
\eta_0\boldsymbol{\dot{\gamma}}
-
\eta_0\tau\overset{\square}{\boldsymbol{\dot{\gamma}}}
+
\eta_0\tau\left(2\epsilon + 1 -a \right)\boldsymbol{\dot{\gamma}}^2.
\end{equation}
We can write this in dimensionless form as
\begin{equation}
\label{gen-SOF-matl-func-tau-dimless}
    \frac{\boldsymbol{\sigma}}{\eta_0/\tau}
=
\tau\boldsymbol{\dot{\gamma}}
-
\tau^2\overset{\square}{\boldsymbol{\dot{\gamma}}}
+
\left(2\epsilon + 1 -a \right)\tau^2\boldsymbol{\dot{\gamma}}^2.
\end{equation}
This representation reveals some important features: 
a characteristic elastic modulus $\eta_0/\tau$, a Weissenberg number $Wi=\tau\boldsymbol{\dot\gamma}$, and a Deborah number that can be associated with a dimensionless time in the derivative from $t^*=t/\tau=De^{-1}$. This is consistent with the G-SOF being an expansion with respect to slow and slowly varying flows, i.e.\ low $Wi$ and low $De$, respectively.

\bibliographystyle{model1-num-names}
\bibliography{references}

\end{document}